# Markov Processes, Hurst Exponents, and Nonlinear Diffusion Equations
*with application to finance*

Kevin E. Bassler, Gemunu H. Gunaratne[+], & Joseph L. McCauley[++]

Physics Department
University of Houston
Houston, Tx. 77204
jmccauley@uh.edu

[+]Institute of Fundamental Studies
Kandy, Sri Lanka

[++]Senior Fellow
**COBERA**
Department of Economics
*J.E.Cairnes Graduate School of Business and Public Policy*
NUI Galway, Ireland

## Abstract

We show by explicit closed form calculations that a Hurst exponent $H \neq 1/2$ does not necessarily imply long time correlations like those found in fractional Brownian motion. We construct a large set of scaling solutions of Fokker-Planck partial differential equations where $H \neq 1/2$. Thus Markov processes, which by construction have no long time correlations, can have $H \neq 1/2$. If a Markov process scales with Hurst exponent $H \neq 1/2$ then it simply means that the process has nonstationary increments. For the scaling solutions, we show how to reduce the calculation of the probability density to a single integration once the diffusion coefficient $D(x,t)$ is specified. As an example, we generate a class of student-t-like densities from the class of quadratic diffusion coefficients. Notably, the Tsallis density is one member of that large class. The Tsallis density is usually thought to result from a nonlinear diffusion equation, but instead we explicitly show that it follows from a Markov process generated by a linear

Fokker-Planck equation, and therefore from a corresponding Langevin equation. Having a Tsallis density with H≠1/2 therefore does not imply dynamics with correlated signals, e.g., like those of fractional Brownian motion. A short review of the requirements for fractional Brownian motion is given for clarity, and we explain why the usual simple argument that H≠1/2 implies correlations fails for Markov processes with scaling solutions. Finally, we discuss the question of scaling of the full Green function g(x,t;x',t') of the Fokker-Planck pde.

## 1. Introduction

Hurst exponents are widely used to characterize stochastic processes, and are often associated with the existence of auto-correlations that describe long term memory in signals [1]. In finance they are used as a measure of the "efficiency" of markets where a value of the Hurst exponent H=1/2 is often said to be required by the efficient market hypothesis (EMH). In this paper we explicitly demonstrate that H≠1/2 is consistent with Markov processes, which by construction have no memory. Therefore, we show that the Hurst exponent alone cannot be used to determine either the existence of long term memory or the efficiency of markets.

As originally defined by Mandelbrot [1], the Hurst exponent H describes (among other things) the scaling of the variance of a stochastic process x(t),

$$\sigma^2 = \int_{-\infty}^{\infty} x^2 f(x,t) dx = ct^{2H} \quad (1)$$

where c is a constant. Here, the initial point in the time series $x_o$=0 is assumed known at t=0. Initially, we limit our analysis to a drift-free process, so that <x>=0.

A Markov process [2,3] is a stochastic process without memory: the conditional probability density for $x(t_{n+1})$ in a time series {$x(t_1)$, $x(t_2)$, ... $x(t_n)$} depends only on $x(t_n)$, and so is independent of all earlier trajectory history $x_1$, ..., $x_{n-1}$. In financial applications our variable x should be understood to be the logarithmic return x(t)=lnp(t)/$p_o$

where p(t) is the price of a financial instrument at time t. A stochastic differential equation

$$dx = \sqrt{D(x,t)}dB(t) \quad (2)$$

describes a Markov process, where B(t) is the Wiener process [4] with <dB>=0, <dB$^2$>=dB$^2$=dt (dB/dt is white noise). That (2) is Markovian is proven very abstractly in the literature [4,5], so we provide the reader with a more straightforward argument in part 3.

Consider next what is required in order to obtain (1), namely, a probability density f(x,t) that scales with a Hurst exponent H, 0<H<1,

$$f(x,t) = t^{-H}F(u); u = x/t^H. \quad (3)$$

The scaling form (3) guarantees that the variance (1) scales. For a Markov process one must also be able to calculate the variance from the sde (2) as

$$\sigma^2 = \int_0^t ds \int_{-\infty}^{\infty} dx f(x,s)D(x,s), \quad (1b)$$

(where we use Ito calculus [4,5]) so that diffusion coefficient scaling

$$D(x,t) = t^{2H-1}D(u), u = x/t^H \quad (4)$$

is a consequence of (3) as well.

The conjecture that H≠1/2 implies correlations follows from the following simple argument [6,7]. Calculate the autocorrelation function in the following way:

$$2\langle \Delta x(t-\Delta t)\Delta x(t+\Delta t)\rangle = \langle (\Delta x(t-\Delta t) + \Delta x(t+\Delta t))^2 \rangle - \langle \Delta x^2(t-\Delta t)\rangle - \langle \Delta x^2(t+\Delta t)\rangle$$

(5)

If the stochastic process x(t) has *stationary increments* [2], requiring for one thing that the mean square fluctuation calculated from any starting point x(t) scales,

$$\langle (x(t+\Delta t) - x(t))^2 \rangle = c\Delta t^{2H}, \quad (6)$$

because it depends only on Δt and not on t, then rescaling the autocorrelation function by the mean square fluctuation, $C(-\Delta t, \Delta t) = \langle \Delta x(t-\Delta t)\Delta x(t+\Delta t)\rangle / \langle \Delta x^2(\Delta t)\rangle$, we have

$$C(-\Delta t, \Delta t) = 2^{2H-1} - 1 \quad (7)$$

so that H≠1/2 implies autocorrelations. This is the likely origin of the common expectation that H≠1/2 violates the condition for both a Markov process and the EMH [6].

However, if (7) would hold for Markov processes then scaling solutions of the form (3,4) could not exist for those processes (2). But we will show in part 4 by direct construction that such solutions do exist, and will show in part 5 that the step from (5) to (7) is doesn't hold for a Markov process with H≠1/2, so that when (5) is calculated directly, then the right hand side vanishes.

This means that an empirical measurement or theoretical prediction of a Hurst exponent, without showing in addition that the increments are stationary or else that the dynamics actually has memory, cannot be interpreted as evidence for autocorrelations (7) in data.

We began this project initially with the aim of understanding where and how the statistics generated by the drift-free nonlinear diffusion equation

$$\frac{\partial f_q}{\partial t} = \frac{1}{2}\frac{\partial^2}{\partial x^2}(f_q^{2-q}) \quad (8)$$

fit into the theory of stochastic processes. The reason for that motivation is that the assumption is made in many papers on Tsallis model dynamics [8,9,10,10b,10c,11] that the nonlinear diffusion eqn. (8) is a "nonlinear Fokker-Planck" partial differential equation (pde) with underlying stochastic differential equation (sde)

$$dx = \sqrt{D(x,t)}dB(t). \quad (2)$$

The standard assumption is that (2) should describe the same process as the solution $f_q(x,t)$ of (8) if we use the diffusion coefficient

$$D_q(x,t) = f_q^{1-q}(x,t) \quad (9)$$

in the sde (2). But the question of consistency arises when one tries to construct a Langevin description (2) for a truly nonlinear diffusion pde (we state the inconsistency explicitly in part 3 below). The question left unanswered in all the papers on the Tsallis model is: what is the density of f(x,t) of x that is generated by (2) if we use (9), where $f_q$ solves the pde (8)? So far, no proof has been given that the two separate methods yield the same probability density in the many papers written while making that assumption. It has also been assumed [8] that the Hurst exponent $H \neq 1/2$ in $f_q$ signals fractional Brownian motion (fBm). We want to analyze these assumptions carefully. Toward that end, we explain in part 3 why (2) is always Markovian. The Tsallis model solution $f_q$ of (8) is exhibited in part 6.

Finally, by studying the one parameter class of quadratic diffusion coefficients, $D(u)=(1+\varepsilon u^2)$ we generate the entire class of student-t-like densities. Student-t distributions have been used frequently in biology and astronomy as well as in finance, so that the diffusion processes analyzed in part 4 will be of interest in those fields. Since the processes are Markovian there is no autocorrelation between different increments $\Delta x$ for nonoverlapping time intervals, but it is well known that there is a form of 'long range dependence' that appears in products of absolute values or squares of random variables [12]. The discussions in [12] have been restricted to near equilibrium stochastic processes (asymptotically stationary processes) like the Ornstein-Uhlenbeck model, and so we plan to discuss the generalization to scaling solutions (3) in a future paper.

Because the term 'stationary' is used differently and is not always clearly defined in some of the recent literature, we next define 'stationary' and 'nonstationary' processes as used in this paper.

## 2. Stationary & Nonstationary Processes, and Processes with Stationary Increments

Here, we define exactly what we mean by a stationary process [2,3,13] and also define a process with stationary increments [2]. Consider a stochastic process x(t) with probability density f(x,t). By stationarity, many authors mean that x(t+Δt)-x(t) depends only on Δt and not on t, but this is the standard definition [2] of a nonstationary process with 'stationary increments'. A stationary process is one where f(x,t)=f(x) is independent of t [2,3,13]. These processes describe statistical equilibrium and steady states because averages of dynamical variables are time-independent. In particular, the mean and variance of a stationary process are constants. An example of an asymptotically stationary process is the Ornstein-Uhlenbeck process

$$dv = -\beta v + \sigma_1 dB(t) \quad (10)$$

describing the approach of the velocity distribution f(v,t) of a Brownian particle to statistical equilibrium (the Maxwell-Boltzmann velocity distribution) [3].

An example of a nonstationary process that generates stationary increments is the Green function

$$g(x,t;x',t') = \frac{1}{\sqrt{4\pi Dt}} e^{-(x-x')^2/2D\Delta t} \quad (11)$$

for the simplest diffusion equation

$$\frac{\partial f}{\partial t} = \frac{D}{2}\frac{\partial^2 f}{\partial x^2} \quad (12)$$

where x is unbounded. Note that (12) also has a (nonnormalizable) stationary solution, f(x)=ax+b, but that that solution is not approached dynamically (is not reached by (11)) as t goes to infinity. If x is instead bounded on both sides (particle in a box), then we get an asymptotically stationary process: f(x,t) approaches a time-independent density f(x) as t goes to infinity [2,3]. Stationary processes are discussed by Heyde and Leonenko [12], who label them "strictly stationary". They also discuss processes with and without stationary increments. Some authors describe a process as 'weakly stationary' if only the mean and variance are constants.

If any moment (mean, variance, etc.) of f(x,t) depends on either t or Δt then the process is by definition nonstationary [2,3,13]. An example is any scaling solution $f(x,t)=t^{-H}F(u)$, $u=x/t^H$, of the Fokker-Planck equation

$$\frac{\partial f(x,t)}{\partial t} = \frac{1}{2}\frac{\partial^2}{\partial x^2}(D(x,t)f(x,t)) \quad (13)$$

Here, the variance $\sigma^2=ct^{2H}$ is always strongly time-dependent, a scaling solution f(x,t) describes a far from equilibrium process. Although F(u) is stationary in u, F(u) obeys no Fokker-Planck equation (correspondingly, the variable u obeys no Langevin equation). The function F(u) is simply the *scale invariant part* of a nonstationary scaling solution f(x,t) of the sde (2).

Nonstationary processes x(t) may have either stationary or nonstationary increments. A nonstationary process with stationary increments is defined by x(t+T)-x(t)=x(T). An example of a nonstationary process with scaling plus stationary increments is Mandelbrot's model of fractional Brownian motion [1], which we exhibit in part 5 below. With stationary increments, not only the variance (1) scales but we obtain the nontrivial condition

$$\langle x((t+T)-x(t))^2 \rangle = cT^{2H}. \quad (14)$$

as well. Nonstationary increments are generated when x(t+T)-x(t)≠x(T), and we also provide an example of this in part 5.

These definitions are precise and differ from the terminology used in some of the modern literature, but are consistent with the dynamical idea of stationary as 'not changing with time'. 'Stationary state' means statistical equilibrium or a constantly driven steady state, so that averages of all dynamical variables are time-independent [2,3,13]. Fluctuations near thermodynamic equilibrium are asymptotically stationary, e.g. For stationary processes, e.g. (10) for larger t, the fluctuation-dissipation theorem [14] holds so that the friction constant β describing regression toward equilibrium in (10) is determined by equilibrium parameters (the temperature of the heat bath).

## 3. Stochastic Differential Equations generate Green Functions for Linear Fokker-Planck PDEs

The proof that the sde (2) is Markovian is not presented transparently in [4,5], so we present a simplified argument here for completeness. Our end result will be the (linear) Fokker-Planck pde

$$\frac{\partial g}{\partial t} = \frac{1}{2}\frac{\partial^2 (Dg)}{\partial x^2} \quad (15)$$

where $D(x,t)$ depends on $(x,t)$ alone, and which describes a Markov process [2,3,4] via it's Green function, $g(x,t;x_o,t_o)$, where $g(x,t;x_o,t)=\delta(x-x_o)$. The Green function is the transition rate density (conditional probability density) for the Markov process. Following Schulten [15], we use Ito calculus to show that a stochastic differential equation, or sde

$$dx = \sqrt{D(x,t)}dB \quad (2)$$

necessarily generates a conditional probability density $g(x,t;x_o,t_o)$ which, by construction, is the Green function for the Fokker-Planck pde (15),

$$g(x,x_o;t,t_o) = \left\langle \delta(\Delta x - \sqrt{D} \bullet \Delta B) \right\rangle = \frac{1}{2\pi}\int_{-\infty}^{\infty} e^{ik\Delta x} \left\langle e^{-i\sqrt{D}\bullet \Delta B} \right\rangle dk. \quad (16)$$

where the dot denotes the Ito product, the stochastic integral of √D(x,t) with respect to B [4]. In all that follows we use Ito calculus because Ito sdes are one to one with (linear) Fokker-Planck pdes, as we will show below. In (16), the average is over all Wiener processes B(t), so that the Green function can be written as a functional integral over locally Gaussian densities with 'local volatility' D(x,t) [16]. By f(x,t) in this paper, we mean the Green function $f(x,t)=g(x,0;t,0)$ for the special case where $x_o=0$ at $t_o=0$. Next, we show why the Fokker-Planck pde (15) is required by the sde (2) (it is well known that the two equations transform one to one under coordinate transformations whenever Ito calculus is used), and why the Fokker-Planck pde *must* be linear. The connection between Wiener integrals for stochastic processes and linear diffusive partial differential equations was first discussed by Kac (see ch. 4 in [17]), but see also Durrett [5] for interesting examples of 'solving a (parabolic) pde by running a Brownian motion (sde)'

Beginning with the sde (2) but with drift included,

$$dx = R(x,t)dt + \sqrt{D(x,t)}dB, \quad (17)$$

consider the time evolution of any dynamical variable A(x) that does not depend explicitly on t (e.g., $A(x)=x^2$). The sde for A is given by Ito's lemma [4,5,15],

$$dA = \left(R\frac{\partial A}{\partial x} + \frac{1}{2}\frac{\partial^2 A}{\partial x^2}\right)dt + \frac{\partial A}{\partial x}\sqrt{D(x,t)}dB. \quad (18)$$

Forming the conditional average

$$\langle dA \rangle = \left(\left\langle R\frac{\partial A}{\partial x}\right\rangle + \left\langle \frac{1}{2}\frac{\partial^2 A}{\partial x^2}\right\rangle\right)dt \quad (19)$$

by using the Green function $g(x,t;x_o,t_o)$ generated by (17) and integrating by parts while ignoring the boundary terms[1], we obtain

$$\int dx A(x)\left[\frac{\partial g}{\partial t} + \frac{\partial(Rg)}{\partial x} - \frac{1}{2}\frac{\partial^2(Dg)}{\partial x^2}\right] = 0, \quad (20)$$

so that for an arbitrary dynamical variable A(x) we get the Fokker-Planck pde

$$\frac{\partial g}{\partial t} = -\frac{\partial(Rg)}{\partial x} + \frac{1}{2}\frac{\partial^2(Dg)}{\partial x^2}. \quad (21)$$

---

[1] If the density g has fat tails, then higher moments will diverge. There, one must be more careful with the boundary terms.

That is, an sde (2) with drift R(x,t) and diffusion D(x,t) generates a (generally nonGaussian) random time series {x(t)} whose histograms at different times t are distributed according to the Green function g of the Markovian linear pde (21), where $g(x,t;x_o,t)=\delta(x-x_o)$. The point here is that Langevin equations generate exactly the same statistics as the corresponding *linear* Fokker-Planck pde.

In what follows we will assume that R may depend on t but not on x, so that we can replace x by x- $\int$Rdt in (21) to get the drift free pde (15). **An x-dependent drift R(x,t) is inconsistent with the scaling form (3).**

At this point we can make a prediction: either the nonlinear pde (8) has no underlying Langevin equation (2), because a nonlinear pde has no Green function and the Green function (transition probability density) is the heart and soul of a Markov process. Or, the pde (8) is really a linear Fokker-Planck pde somehow disguised as a nonlinear diffusion equation.

## 4. Markov Processes with Scaling Solutions

Until the last section of this paper, we restrict to the case where g(x,0,;t,0)=f(x,t) because we will show explicitly below that only these solutions exhibit exact scaling properties (1), (3), (4), and also because the density f(x,t) is what one observes directly in histograms of finance market returns data. The full Green function g(x,x';t,t') is needed in principle for exact option pricing (see [16,21] for another requirement) but cannot be calculated in closed analytic form when D(x,t) depends on both x and t and scaling (3, 4) holds. If f and therefore D scale according to (3) and (4), then the variance scales exactly as (1) with Hurst exponent H. The empirical evidence for the data collapse predicted by (3) will be presented in a separate paper on financial markets [18]. The question of scaling of the full Green function is discussed at the end of this paper.

Inserting the scaling forms (3) and (4) into the pde (15), we obtain[2]

---

[2] We emphasize that the drift has been subtracted out of the pde (21) to yield (15). This requires that R(x,t) is independent of x. A x-independent drift R(t) is absolutely necessary for the scaling forms (3,4).

$$2H(uF(u))' + (D(u)F(u))'' = 0 \quad (22)$$

whose solution is given by

$$F(u) = \frac{C}{D(u)} e^{-2H \int u du / D(u)}. \quad (23)$$

Note that (23) describes the scale invariant part of *nonstationary* solutions f(x,t) (3). This generalizes our earlier results [19,20,20b] to include H≠1/2.

Next, we study the class of quadratic diffusion coefficients

$$D(u) = d'(\varepsilon)(1 + \varepsilon u^2), \quad (24)$$

which yields the two parameter (ε,H) class of student-t- like densities

$$F(u) = C'(1 + \varepsilon u^2)^{-1 - H/\varepsilon d'(\varepsilon)} \quad (25)$$

with tail exponent μ=2+2H/εd'(ε), and where H and ε are independent parameters to be determined empirically. This shows where student-t-like distributions come from. With d'(ε)=1 we obtain the generalization of our earlier prediction [20] to arbitrary H, 0<H<1, with μ=2+2H/ε. Here, we generate all fat tail exponents 2<μ<∞, and obtain a finite variance (1) scaling as $\sigma^2 = ct^{2H}$ whenever μ>3 (for 2≤μ≤3 the variance is infinite). For large u this model fits the fat tails in financial data for all times t [18]. For small to moderate returns finance market histograms are approximately exponentially distributed [16,21], and a complete empirical analysis of market data will be presented in a forthcoming separate paper [18].

## 5. What does H mean?

The assumption that "H≠1/2 implies long time correlations" is often found in the literature (see, however, Mandelbrot [22]). An example where H≠1/2 but there are no autocorrelations is the Levy density

[23]. In that case, H describes the scaling of the peak of the density and also the tail exponent. For the Levy densities the variance is infinite.

Consider the Markov process (2) in its integrated form

$$\Delta x(t, t + \Delta t) = \int_t^{t+\Delta t} \sqrt{D(x(s), s)} dB(s) \quad (26)$$

Even when scaling (3,4) holds with $H \neq 1/2$, then on quite general grounds there can be no autocorrelation in the increments $\Delta x(t, t-\Delta t)$, $\Delta x(t, t+\Delta t)$ over two nonoverrlapping time intervals

$$[t, t + \Delta t] \cap [t - \Delta t, t] = \varnothing. \quad (27)$$

This is easy to see: by definition of the Ito integral:

$$\langle \Delta x(t, t - \Delta t) \Delta x(t, t + \Delta t) \rangle = \int_t^{t+\Delta t} ds \int_{t-\Delta t}^t dw \langle D(x(w), w) D(x(s), s) \rangle \langle dB(w) dB(s) \rangle = 0$$
(28)

because <dB(w)dB(s)>=0 for nonoverlapping time intervals dw and ds [4,5,15]. The function D(x,t) is called 'nonanticipating' [4]. This just means that, by Ito's definition of the stochastic integral (26), the function D(x,t) of random variable x and the random increment dB(t) from t to t+dt are statistically dependent because x(t) was determined in the sde (2) by the Wiener increment dB(t-dt) before dB(t) occurs. That is, D(x(t),t) cannot 'anticipate' the next random increment dB(t) in (26).

The passage from (5) to (7) requires a usually unstated assumption of stationary increments. If the nonstationary stochastic process x(t) has *nonstationary increments*, requiring that the mean square fluctuation about x(t) depends both on $\Delta t$ and t, then the passage from (5) to (7) is not possible. The argument that $H \neq 1/2$ implies long time correlations fails for Markov processes precisely because the stochastic integral (26) with the scaling forms (3,4) describes a nonstationary process with *nonstationary increments* whenever $H \neq 1/2$. Only for $H=1/2$ do we retrieve a nonstationary process with stationary increments. When $H \neq 1/2$ then (26) combined with scaling (4) yields (with x(0)=0)

$$x(t+T) - x(t) = \int_0^{t+T} \sqrt{D(x(s),s)}\, dB(s) - \int_0^t \sqrt{D(x(s),s)}\, dB(s)$$

$$= \int_t^{t+T} \sqrt{D(x(s),s)}\, dB(s) = \int_t^{t+T} |s|^{H-1/2} \sqrt{D(u)}\, dB(s)$$

$$= \int_0^T |s+t|^{H-1/2} \sqrt{D(x/|s+t|^H)}\, dB(s+t) \neq x(T) \tag{26b}$$

whereas we retrieve stationary increments $x(t+T)-x(t)=x(T)$ for $H=1/2$ with probability one, e.g., with $H=1/2$ we find that $<(x(t+T)-x(t))^2>=<x^2(T)>=cT$. Furthermore, direct calculation of the autocorrelation formulated as (5) shows that the right hand side of (5) vanishes independently of the value of H, in agreement with (28) above.

We've seen above that a Hurst exponent $H \neq 1/2$ is consistent with a Markov process. One only needs the scaling forms (3,4), and the Fokker-Planck pde (15) is then satisfied by $f(x,t)=t^{-H}F(u)$ with $u=x/t^H$ where $F(u)$ is given by (23). This Hurst exponent does not imply long time correlations, so what does $H \neq 1/2$ mean? The appearance of $H \neq 1/2$ in a Markov process signals underlying dynamics with *nonstationary increments*, and this knowledge should be useful for data analysis.

From a purely theoretical standpoint, a Hurst exponent $H \neq 1/2$ for a scale free Markov process can be eliminated by a change of time variable (a corollary is that any Markov process with $H=1/2$ can be converted superficially into one with $H \neq 1/2$ by a change of time scale). Note that for any diffusion coefficient of the form $D(x,t)=h(t)d(x,t)$, the prefactor $h(t)$ can always be absorbed into a redefinition of the time scale in the drift-free Fokker-Planck pde (15), $d\tau=h(t)dt$. Likewise, with the choice of time variable $\tau=t^H$, the pde (15) with the scaling forms (2) and (1) always yields $\sigma^2=c\tau$. So a drift free Markov process with nonstationary increments can be transformed formally into one with stationary increments by the appropriate change of time scale, and vice-versa.

*There can be no correlations for nonoverlapping time intervals because (26) is Markovian*, whether $H=1/2$ or $H \neq 1/2$ plays no role. This is why Markov dynamics reflect the EMH: a Markovian market is impossible to beat. Real markets are very hard to beat systematically over the long haul, so that a Markov model provides us with a very good zeroth order approximation to real financial markets. Another way to say it is, with drift subtracted out, a market is pure (nonGaussian) noise, in agreement with Black's idea of the importance of 'noise traders' [24]. When $H \neq 1/2$ combined with stationary increments in x(t) then there is either persistence or antipersistence of autocorrelations for nonoverlapping time intervals, as in fractional Brownian motion [1]. Fractional Brownian motion (fBm) is inherently nonMarkovian. In principle, a market with $H \neq 1/2$ plus stationary increments has correlations that may be exploited for profit, so that such a market is not "efficient".

One can construct models of fractional Brownian motion as follows. With $k(t,s)=t^{H-1/2}K(u)$, $u=t/s$, a stochastic process of the form

$$x_H(t) = \int_{t_o}^{t} k(t,s)dB(s) \qquad (29)$$

generates long time autocorrelations for nonoverlapping time intervals but doesn't scale. Scaling is obtained iff. $t_o=0$ or $-\infty$. For the former case the increments of (29) are not stationary, but one may obtain stationary increments for $t_o = -\infty$, depending on the form of the function k(t,s). In that case, we have the scaling law $\sigma^2=<x^2>=ct^{2H}$. If the kernel k(t,s) is such that $x_H(t)$ has stationary increments [1],

$$x_H(t+T) - x_H(t) = \int_{-\infty}^{T} k(T,s)dB(s) = x_H(T), \qquad (30)$$

then a simple prediction (a generalization of (7)) for the autocorrlelations of fBm over nonoverlapping time intervals follows: with the autocorrelation function defined more generally by

$$C(S_1, S_2) = \langle (x_H(t+\Delta t_1) - x_H(t))(x_H(-t) - x_H(-t-\Delta t_2)) \rangle / \sigma_1^2 \sigma_2^2 \qquad (31)$$

where $S_1=\Delta t_1/t$, $S_2=\Delta t_2/t$, we obtain [1]

$$C(S_1, S_2) = [(1+S_1+S_2)^{2H} + 1 - (1+S_1)^{2H} - (1+S_2)^{2H}]/2(S_1 S_2)^H \qquad (32)$$

This prediction can easily be generalized to allow widely separated time intervals $[t_1-\Delta t_1, t_2+\Delta t_2)$ where $t_1<t_2$. Mandelbrot [1] has provided us with an illuminating example of fBm with stationary increments (25),

$$x_H(t) = \int_{-\infty}^{0}[(t-s)^{H-1/2} - (-s)^{H-1/2}]dB(s) + \int_{0}^{t}[(t-s)^{H-1/2}dB(s)$$

(29b)

Clearly, such long time correlations are nonMarkovian and violate the EMH. Note that the correlations (32) vanish if $H=1/2$, whereas in a Markov process the correlations vanish for all values of H.

Contrary to statements [6,7] in the literature, (29b) is not a Gaussian process. With $D(u)=1$ (23) becomes $F(u)=\exp(-u^2)$ with $u=x/t^H$, but this solution of (15) describes a Markov process with no autocorrelations at all. One cant rightly state that (29b) arises from Gaussian increments $dB(t)$, but then so does every other stochastic process described by an Ito integral, and those processes are typically far from Gaussian distributed, as is (29b). We do not yet know the functional form $f(x_H,t)$ of the density of (29b), other than that it must scale like (3) and cannot be Markovian.

A Markov process provides a sufficient but not necessary condition for the EMH. Since finance market data can be described as approximately Markovian [16,19], to zeroth order, then searching for fBm or other memory in market data would be a search for a way to make small profit margins by placing big bets.

## 6. The Tsallis Density

It is easy to check by direct calculation that a normalized solution of (8) is given self-consistently by

$$f_q(x,t) = (c(2-q)(3-q))^{-H} t^{-H} (1 + (q-1)x^2/C^2(q)t^{2H})^{1/(1-q)} \quad (33)$$

with $H=1/(3-q)$, where

$$C(q) = c^{(q-1)/2(3-q)}((2-q)(3-q))^H \quad (34)$$

and

$$c^{1/2} = \int_{-\infty}^{\infty} du(1+(q-1)u^2)^{1/(1-q)} \quad (35)$$

is the normalization constant [11]. Normalization is not overdetermined because the pde (8) satisfies probability conservation. The fat tail exponent, $f(x,t) \approx x^{-\mu}$ for $x \gg 1$, is $\mu = 2/(q-1)$. This model has the constraint that the tail exponent $\mu$ is fixed by the Hurst exponent H, or vice-versa. E.g., if $H = 1/2$, then there are no fat tails, the density is Gaussian.

Inserting (33) into (9) yields the diffusion coefficient

$$D_q(x,t) = (c(2-q)(3-q))^{2H-1} t^{2H-1}(1+(q-1)x^2/C^2(q)t^{2H}) \quad (36)$$

which we conveniently rewrite as

$$D_q(x,t) = d(q)t^{2H-1}(1+((q-1)/C^2(q))u^2) = t^{2H-1}D_q(u) \quad (37)$$

To compare (33) with (25), we need only write $\varepsilon = (q-1)/C^2(q)$ and $d'(\varepsilon) = d(q)$. Our Fokker-Planck-generated density $f(x,t)$ given by (25) reduces exactly to (33) when $H = 1/(3-q)$. This means that $f_q$ actually satisfies the *linear* Fokker-Planck pde

$$\frac{\partial f}{\partial t} = \frac{1}{2}\frac{\partial^2(D_q f)}{\partial x^2} \quad (8b)$$

and so (8), for the Tsallis solution (33), is really a linear pde disguised as a nonlinear one.

A nonlinear disguise is possible for our entire two-parameter student-t-like class solutions (25), because for quadratic diffusion (24), $D(u) = d'(\varepsilon)(1+\varepsilon u^2)$, the solution of the Fokker-Planck pde (8) is a power of the diffusion coefficient, $F(u) = CD(u)^{-1-H/\varepsilon d'(\varepsilon)}$. All of these solutions trivially satisfy a modified form of the nonlinear pde (8), but rewriting (8b) as a nonlinear pde in the case of quadratic diffusion superficially masks the Markovian nature of Tsallis dynamics.

The claim made is in Borland [8] and elsewhere that Tsallis model (8) generates fractional Brownian motion, but this is not correct. The Tsallis density (33) is Markovian and so cannot describe long-time correlated signals like fBm. There, $H=1/(3-q)\neq 1/2$ merely signals that the increments x(t) are nonstationary.

In a Langevin/Fokker-Planck approach with x-dependent drift, Kaniadakis and Lapenta [9] did not reproduce the time dependence of the Tsallis density (33) with $H=1/(3-q)$. In their formulation using an x-dependent drift term in the Fokker-Planck pde, they find a time-dependent solution that does not scale with a Hurst exponent H. That is, nonscaling solutions are certainly possible. And as we have pointed out, scaling of f(x,t) is impossible when the drift depends on x.

But what about truly nonlinear diffusion? The linear pde (8b) solves a unique initial value problem, and unique boundary value problems as well. But we do not know if the nonlinear pde

$$\frac{\partial f}{\partial t} = \frac{1}{2}\frac{\partial^2}{\partial x^2}(f^m) \quad (8c)$$

with $m\neq 1$ has a unique solution for a specified initial condition f(x,0). There may be solutions other than the trivial self-consistent solution (33), and there we cannot rule out the possibility of long time memory in (8c).

For a discussion of the general properties of nonMarkovian linear pdes with memory, see [25]. See also Hillerbrand and Friedrich [25b] for nonMarkov densities of the form $f(x,t)=t^{-3/2}F(x/t^{1/2})$ based on memory in the diffusion coefficient.

## 7. Scaling and the Green function

Finally, a few words about the full Green function of the Fokker-Planck pde (15). So far, we've restricted to a special case where

f(x,t)=g(x,t;0,0). In this case, as we've shown by direct construction, the scaling (1,3,4) is exact. For the general Green function g(x,t;x',t') with x'≠0 scaling is not exact and may not exist at all.

If we assume that g(x,t;x',t')=g(x,x';Δt), and if we in addition make the (unproven) scaling Ansatz

$$g(x,x';\Delta t) = \Delta t^{-H} G(u,u_o) \quad (39)$$

where $u=x/t^H$, $u_o=x_o/t^H$, then we would have a mean square fluctuation

$$\langle (x-x_o)^2 \rangle = \int_t^{t+\Delta t} ds \Delta s^{2H-1} \int_{-\infty}^{\infty} du (u-u_o)^2 G(u,u_o) \quad (40)$$

with Δs=s-t. This doesn't yield a simple expression for nonstationary increments unless $G(u,u_o)=G(u-u_o)$, because $u_o=x_o/\Delta s$. We can offer no theoretical prediction for the Green function when x'≠0.

In a future paper we will analyze option pricing and the construction of option prices as Martingales, both from the standpoint of stochastic differential equations [26] and generalized Black-Scholes equations [16]. A key observation in that case is that, with fat tails, the option price diverges in the continuum market theory [20b,27]. This result differs markedly from the finite option prices predicted in [11,28].

## 8. Summary and Conclusions

Hurst exponents H≠1/2 are perfectly consistent with Markov processes and the EMH. A Hurst exponent, taken alone, tells us nothing about autocorrelations. Scaling solutions with arbitrary Hurst exponents H can be reduced for Markov processes to a single integration. A truly nonlinear diffusion equation has no underlying Langevin description. Any nonlinear diffusion equation with a Langevin description is a linear Fokker-Planck equation in disguised form. The Tsallis model is Markovian, does not describe fractional Brownian motion. A Hurst exponent H≠1/2 in a Markov process x(t) describes nonstationary increments, not autocorrelations in x(t).

## Acknowledgement

KEB is supported by the NSF through grants #DMR-0406323 and #DMR-0427938, by SI International and the AFRL, and by TcSUH. GHG is supported by the NSF through grant #PHY-0201001 and by TcSUH. JMC thanks C. Küffner for reading the manuscript and suggesting changes that made it less redundant and more readable, and also to Harry Thomas for a careful and critical reading. We're grateful to a referee for pointing out references 10b and 10c, and to R. Friedrich for sending us preprint 25b.